\documentclass{article}
\usepackage{spconf,amsmath,graphicx,enumitem}
\usepackage{cite,subfigure,amssymb,tablefootnote,multirow,makecell}
\usepackage[colorlinks,
            linkcolor=red,
            anchorcolor=blue,
            citecolor=green
            ]{hyperref}
\title{UFO2: A unified pre-training framework for online and offline speech recognition}
\name{Li Fu, Siqi Li, Qingtao Li, Liping Deng, Fangzhu Li, Lu Fan, Meng Chen, Xiaodong He}
\address{JD AI Research, Beijing, China}

\begin{document}
\ninept
\maketitle
\begin{abstract}
In this paper, we propose a \underline{U}nified pre-training \underline{F}ramework for \underline{O}nline and \underline{O}ffline (UFO2) Automatic Speech Recognition (ASR), which 1) simplifies the two separate training workflows for online and offline modes into one process, and 2) improves the Word Error Rate (WER) performance with limited utterance annotating. Specifically, we extend the conventional offline-mode Self-Supervised Learning (SSL)-based ASR approach to a unified manner, where the model training is conditioned on both the full-context and dynamic-chunked inputs. To enhance the pre-trained representation model, stop-gradient operation is applied to decouple the online-mode objectives to the quantizer. Moreover, in both the pre-training and the downstream fine-tuning stages, joint losses are proposed to train the unified model with full-weight sharing for the two modes. Experimental results on the LibriSpeech dataset show that UFO2 outperforms the SSL-based baseline method by 29.7\% and 18.2\% relative WER reduction in offline and online modes, respectively.
\end{abstract}
\begin{keywords}
Automatic speech recognition, self-supervised learning, online and offline unified model
\end{keywords}
\vspace{-0.04in}
\section{Introduction}
\label{sec:intro}
\vspace{-0.04in}
In recent years, Self-Supervised Learning (SSL) has received much attention in the Automatic Speech Recognition (ASR) domain~\cite{baevski2020wav2vec,hsu2021hubert,chen2022wavlm,yang2021superb,ravanelli2020multi,babu2021xls,gao2022self}. Generally, the SSL-based ASR approach first pre-trains a speech representation encoder on numerous unlabeled utterances via self-supervised strategies (e.g. masking, quantization, contrasting~\cite{baevski2020wav2vec}), and then fine-tunes the model on labeled data with ASR objectives. It has shown great potential in improving the ASR performance with limited speech labeling, which is very promising and valuable when the human-annotated utterances are expensive or scarce~\cite{mohamed2022self}.

Based on how the tokens are emitted, ASR systems are typically categorized by their use in: 1) {\it online mode} (a.k.a. streaming), which is developed to emit each hypothesized word as quickly and accurately as possible when it is spoken~\cite{miao2020transformer,he2019streaming,li2020towards,doutre2021improving}, and 2) {\it offline mode} (a.k.a. non-streaming), which aims to accurately emit the complete hypotheses after processing a full utterance~\cite{li2020popular,fu2021incremental,braun2020gpu}. However, most existing SSL-based ASR methods focus on the pre-training in an offline manner, i.e. each represented feature is conditioned on the full-context inputs~\cite{karimi2022deploying}. As for the downstream online ASR model that no (or limited) future context is permitted, the accuracy performance might be hindered due to the mode inconsistency between the pre-training and fine-tuning~\cite{chiu2022self}. One might pre-train the representation encoder in online manners, while the model might suffer a heavy burden on the representation learning when a large proportion of the utterance (i.e. future context) is unavailable~\cite{fu2021scala}.

Different from the offline-mode SSL, there are only few works about pre-training for online ASR models. Chiu et al.~\cite{chiu2022self} explored replacing the learnable quantizer in~\cite{baevski2020wav2vec} by a Random-projection Quantizer (RQ), to make the quantized representation independent to the recognition mode. Although the RQ method was separately evaluated on online and offline models with 0.6 billion parameters, as mentioned in~\cite{chiu2022self} the random strategy of the quantizer and smaller model sizes that are more efficient for online ASR tasks still need to be further investigated. Cao et al.~\cite{cao2021improving} trained an SSL-based offline ASR model (teacher model), and then adopted knowledge distillation~\cite{hinton2015distilling} to guide the fine-tuning of an online model (student model). Nevertheless, besides introducing the additional offline model optimization and distillation strategies under the SSL framework, the online model to be fine-tuned was still initialized from an offline-mode representation encoder. Moreover, in the existing SSL-based works, the online and offline ASR systems were processed separately, causing high costs in model development and training workflows for applications in different modes~\cite{yao2021wenet,yu2020dual,liu2022learning}.

\begin{figure}[t]
  \centering
  \includegraphics[width=0.48\textwidth]{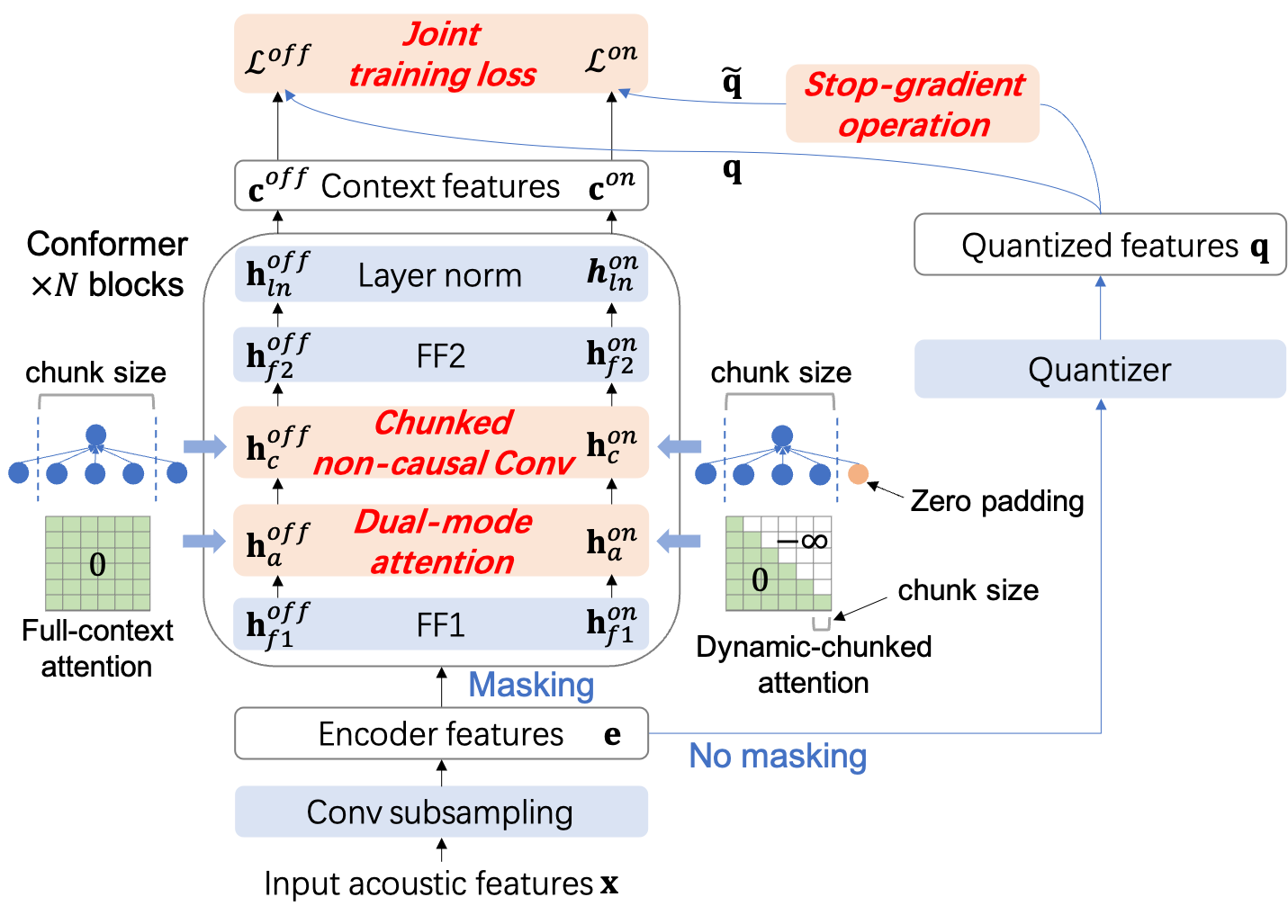}
  \vspace{-0.3in}
  \caption{Overview of the proposed UFO2: Encoder features $\mathbf e$ are masked and fed to Conformer blocks to extract offline/online latent features ($\mathbf{h}_*^{off}$/$\mathbf{h}_*^{on}$) conditioned on full-context/dynamic-chunked inputs. Then context features $\mathbf{c}^{off}$/$\mathbf{c}^{on}$ and quantized features ${\mathbf{q}}$/$\widetilde{\mathbf{q}}$ (w/ stop gradient) are used for a joint of losses $\mathcal{L}^{off}$ and $\mathcal{L}^{on}$.}
  \label{fig1}
  \vspace{-0.05in}
\end{figure}

\begin{table*}[t]
    \vspace{-0.2in}
	\centering  
	\caption{System performance in WER ($\%$) (best performance is marked in bold).}
	\label{table1}  
	\resizebox{0.75\linewidth}{!}{
	\begin{tabular}{c|l|c|c|c|c|c|c|c|c|c|c}  
		\hline  
		~&\multirow{3}{*}{Methods} & \multicolumn{2}{c|}{Training dataset} & \multicolumn{4}{c|}{test clean} & \multicolumn{4}{c}{test other} \\
		\cline{3-12}
		~&~&Unlabeled&Labeled&\multicolumn{2}{c|}{offline} & \multicolumn{2}{c|}{online} & \multicolumn{2}{c|}{offline} & \multicolumn{2}{c}{online} \\
		\cline{3-12}
		~&~&$\boldsymbol{D}_u$&$\boldsymbol{D}_l$&CPBS&AR&CPBS&AR&CPBS&AR&CPBS&AR\\
		\hline
		\hline
		S1&Conformer-offline~\cite{gulati2020conformer} & NA & 100h & 8.82 & 8.11 & NA & NA & 23.33 & 22.19 & NA & NA \\
		S2&Wav2vec2-base~\cite{baevski2020wav2vec} & 960h & 100h & 6.10 & NA & NA & NA & 13.30 & NA & NA & NA  \\
		S3&Wav2vec2-Conformer~\cite{zhang2020pushing} & 960h & 100h & 5.80 & 5.49 & NA & NA & 13.21 & 12.50 & NA & NA  \\
		\hline
		S4&Conformer-unified~\cite{yao2021wenet} & NA & 100h & 10.64 & 9.83 & 11.39 & 10.53 & 26.58 & 25.55 & 28.40 & 27.34  \\
		S5&Wav2vec2-Conformer-unified & 960h & 100h & 7.38 & 6.78 & 8.27& 7.66 & 18.47 & 17.65 & 21.08 & 20.31  \\
		S6&UFO2 & 960h & 100h & {\bf 5.46} & {\bf 4.99} & {\bf 7.06} & {\bf 6.35} & {\bf 12.38} & {\bf 11.75} & {\bf 16.87} & {\bf 16.05}  \\
		\hline
	\end{tabular}
	}
	\vspace{-0.19in}
\end{table*}

In this paper, a novel \underline{U}nified pre-training \underline{F}ramework is proposed to improve the speech representation learning for downstream \underline{O}nline and \underline{O}ffline (UFO2) ASR tasks. In particular, UFO2 simplifies the training workflows by unifying the online and offline modes into a single model. As shown in Fig.~\ref{fig1}, different from the most representative SSL-based approach -- Wav2vec2~\cite{baevski2020wav2vec} and its variants, e.g. Wav2vec2-Conformer~\cite{zhang2020pushing}, we extend the offline-mode approach to a unified manner via four strategies on the feature extraction and training objectives. {\bf 1) Dual-mode attention}. To train a unified representation encoder, the full-context Multi-Headed Self-Attention (MHSA) in the Conformer block~\cite{gulati2020conformer} is used to extract offline-mode features conditioned on the complete utterance. Simultaneously, the dynamic-chunked MHSA~\cite{yao2021wenet} is adopted to mimic different latency ranges for online-mode learning. {\bf 2) Chunked non-causal Convolution (Conv)}. Instead of using the popular causal-Conv Conformer to build a unified model~\cite{yao2021wenet}, we leverage non-causal Convs to enhance the local feature extraction for online mode, while the positions in the future chunks are padded with zero to make the latency strictly controlled within the chunk size. {\bf 3) Stop-gradient operation.} The online and offline representation models share all the encoder and quantizer weights. However, to further improve the representation, stop gradient is operated to decouple the impact of the online-mode objectives to the quantizer. {\bf 4) Joint training loss.} The online and offline objectives are aggregated to train the unified model, which encourages the two modes to promote each other. Experimental results show that UFO2 achieves obvious accuracy improvements in comparison with the baseline methods in both the online and offline modes. Our main contributions are summarized as follows: 

\vspace{-0.05in}
\begin{itemize}[leftmargin=0.12in]
	\setlength{\itemsep}{-2pt}
	\item To the best of our knowledge, this is the first work exploring a unified pre-training framework for online and offline ASR.
	\item We propose a new SSL-based ASR approach, named UFO2, by introducing simple and effective strategies for unified training.
	\item We show the effectiveness of our method on the LibriSpeech 100 hours experiments with significantly lower Word Error Rates (WERs) compared with the baseline methods.
\end{itemize}

\vspace{-0.1in}
\section{Related work}
\vspace{-0.05in}
\label{sec:related_work}

{\bf Unified ASR in supervised learning.} Recently, it has been shown favorable to unify online and offline ASR systems. Conformer-unified~\cite{yao2021wenet} randomly sampled one from the two modes when training a single causal-Conv Conformer model. To improve the model quality, the online and offline ASR losses were joint for dual-mode learning with weight sharing~\cite{yu2020dual}: the non-causal Conv in the Conformer was used for offline feature extraction, while the Conv's right-half weights were masked to mimic a causal Conv for online mode. Inspired by~\cite{yu2020dual}, a self-pruning method was proposed to unify a compact sparse on-device online ASR model and a large dense offline model~\cite{liu2022learning}. The existing unified systems are mainly discussed in the supervised setting, while we focus on pre-training the unified model on unlabeled utterances. Note that the strategies in our method could also be extended for the supervised ASR task.

\noindent{\bf SSL-based ASR.} To overcome the need for labeled training data, SSL-based ASR has been a hot topic in recent years. Wav2vec~\cite{Schneider2019wav2vec} adopted contrastive learning to train speech representations and improved the ASR accuracy. To enhance the performance, a quantizer was designed in Vq-Wav2vec~\cite{Baevski2020vqwav2vec} to learn discrete speech representations. A further performance gain was achieved in Wav2vec2~\cite{baevski2020wav2vec} by combining masking, quantizer, and contrasting strategies in the pre-training. To optimize the model architecture, Wav2vec2-Conformer~\cite{zhang2020pushing} was proposed by replacing the Transformer block of Wav2vec2 with the Conformer block. More work about the SSL-based ASR approach can be referred to~\cite{mohamed2022self}. However, different from the existing works that assume an offline-mode setting~\cite{baevski2020wav2vec,zhang2020pushing} or optimize the offline and online models separately~\cite{cao2021improving,chiu2022self}, our UFO2 unifies online and offline modes in both the pre-training and fine-tuning stages to simplify the training workflows, and encourages the two modes to promote mutually during the training.

\section{Our proposed approach}
\label{sec:method}
\vspace{-0.05in}
\subsection{Problem formulation}
\vspace{-0.05in}

{\bf Unlabeled dataset:} $\boldsymbol{D}_u=\{{\mathbf x}^i|i\in{\{1,\cdots,N_u\}}\}$ with $N_u$ the number of utterances; $\mathbf{x}^{i}\in\mathbf{R}^{{F}\times{T}_{i}}$ is the $i^{th}$ sample which is a sequence of $F$-dimensional acoustic features with length ${T}_{i}$.\\
{\bf Labeled dataset:} $\boldsymbol{D}_l=\{{\mathbf x}_l^i,{\mathbf y}_l^i|i\in{\{1,\cdots,N_l\}}\}$ with ${N_l}$ the number of labeled samples; $\mathbf{y}_l^i\in\mathbf{L}^{{U}_{i}}$ is the label sequence (with length ${U}_{i}$) of utterance ${\mathbf x}_l^i$, where $\mathbf{L}$ is the finite label character.\\
{\bf Aim of UFO2:} Leverage the unlabeled dataset $\boldsymbol{D}_u$ to learn speech representations and then fine-tune the model on $\boldsymbol{D}_l$ to obtain a single ASR model that performs well in both the online and offline modes.

\begin{table*}[t]
    \vspace{-0.2in}
	\centering  
	\caption{Ablation studies with different training losses and model choices in WER ($\%$).}
	\label{table2}  
	\resizebox{0.879\linewidth}{!}{
	\begin{tabular}{c|c|c|c|c|c|c|c|c|c|c|c}  
		\hline  
		~&\ Pre-training & Conv in Conformer ASR& Fine-tuning & \multicolumn{4}{c|}{test clean} & \multicolumn{4}{c}{test other} \\
		\cline{2-12}
		~&$\triangle$: offline loss&$\triangle$: causal&$\triangle$: random-mode loss&\multicolumn{2}{c|}{offline} & \multicolumn{2}{c|}{online} & \multicolumn{2}{c|}{offline} & \multicolumn{2}{c}{online} \\
		\cline{5-12}
		~&\checkmark: joint loss&\checkmark: chunked non-causal&\checkmark: joint loss&CPBS&AR&CPBS&AR&CPBS&AR&CPBS&AR\\
		\hline
		\hline
		A1&$\triangle$ & $\triangle$ & $\triangle$ & 7.38&6.78&8.27&7.66&18.47&17.65&21.08&20.31\\
		A2&$\triangle$ & $\triangle$ & \checkmark & 6.15&5.65&8.10&7.57&14.39&13.62&19.71&18.87 \\
		A3&$\triangle$ & \checkmark & $\triangle$ & 6.11&5.62&7.71&7.05&14.36&13.62&18.96&18.11 \\
		A4&$\triangle$ & \checkmark & \checkmark & 5.85&5.45&7.63&7.08&14.16&13.39&18.82&18.13  \\
		A5&\checkmark & \checkmark & \checkmark & {\bf 5.46} & {\bf 4.99} & {\bf 7.06} & {\bf 6.35} & {\bf 12.38} & {\bf 11.75} & {\bf 16.87} & {\bf 16.05}  \\
		\hline
	\end{tabular}
	}
	\vspace{-0.2in}
\end{table*}

\vspace{-0.15in}
\subsection{Model architecture}
\label{sec:model_architecture}
\vspace{-0.05in}

Our model mainly contains a Conv subsampling encoder, a Conformer context encoder, and a quantizer module, as shown in Fig.~\ref{fig1}. Given input acoustic features ${\mathbf x}$, the Conv subsampling encoder outputs encoder features $\mathbf{e}$ with a $4\times$ reduction in the sequence length, which are then processed in the following two ways. 1) The encoder features are masked on sampled 6.5\% time-steps and the 10 consecutive steps~\cite{baevski2020wav2vec}, and then fed to a stack of $N$ Conformer blocks to obtain offline and online context features $\mathbf{c}^{off}$ and $\mathbf{c}^{on}$ simultaneously. Each Conformer block contains two Feed Forward (FF1, FF2) modules sandwiching the MHSA and non-causal Conv modules~\cite{gulati2020conformer}, where the offline and online latent features ($\mathbf{h}_*^{off}$, $\mathbf{h}_*^{on}$) are extracted to adapt the model for the two modes. 2) The encoder features $\mathbf{e}$ (without masking) are quantized to features $\mathbf{q}$ selected from $G$ codebooks with $V$ entry vectors via the quantizer module~\cite{baevski2020wav2vec}. Finally, the model is pre-trained with a self-supervised contrastive loss, which is applied to distinguish the quantized feature at the same time-step of each masked context feature from other masked samples~\cite{baevski2020wav2vec}.

\vspace{-0.139in}
\subsection{Strategies for unified model}
\label{sec:training}
\vspace{-0.05in}
To enhance the performance of the unified model, four strategies (highlighted in red, in Fig.~\ref{fig1}) on the feature extraction and training objectives are proposed as follows.

\noindent {\bf 1) Dual-mode attention.} One of the key parts for unifying the online and offline modes is to enable the model extracting dual-mode features~\cite{yu2020dual}. In the proposed UFO2, a bias matrix $\mathbf B$ (unlearnable) is added to the attention logits~\cite{vaswani2017attention} $\mathbf A$ of the MHSA module, yields the dual-mode attention matrix ${\mathbf A}_d={\rm Softmax}({\mathbf A}+{\mathbf B})$ as below.
\begin{itemize}[leftmargin=0.12in]
	\setlength{\itemsep}{-2pt}
	\item {\it Full-context attention:} $\mathbf B$ is set to a zero matrix to output the normal attention conditioned on the full utterance for offline mode.
	\item {\it Dynamic-chunked attention:} We first segment the inputs into non-overlapping chunks with size $c_s$, and then set the top-right elements of $\mathbf B$ to $-\infty$. The $-\infty$ values lead to zero attention scores via the Softmax function, i.e. each online-mode feature cannot attend to frames belonging to the future chunks~\cite{chen2021developing}. To enhance the online-mode learning, $c_s$ is dynamically selected for each training mini-batch to mimic latency ranges from 0 to 1 second~\cite{yao2021wenet}.
\end{itemize}

\begin{figure}[ht]
  \centering
  \vspace{-0.12in}
  \includegraphics[width=0.47\textwidth]{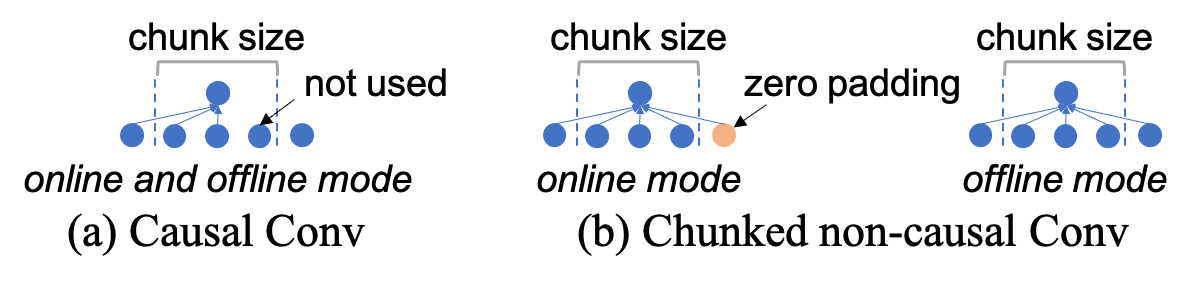}
  \vspace{-0.15in}
  \caption{Advantages of chunked non-causal Conv, over causal Conv, in leveraging more local features in online and offline modes.}
  \label{fig2}
  \vspace{-0.12in}
\end{figure}

\noindent {\bf 2) Chunked non-causal Conv.} Typically, causal-Conv~\cite{yao2021wenet} is used for Conformer-based online ASR models to extract local features being independent on its future inputs. It mitigates the latency increasing problem when stacking multiple non-causal-Conv Conformer blocks~\cite{shi2022streaming}. However, the causal-Conv may hinder the dynamic-chunked approach since the future inputs belonging to the current chunk are not used (see Fig.~\ref{fig2}). To address this issue, we design a chunked non-causal Conv for the unified model. In online mode, the positions in the future chunks are padded with zero to ensure the latency strictly controlled within the chunk size; the Conv's weights are fully shared to offline mode without chunk restriction. Compared with the causal Conv, the main advantages of the chunked non-causal Conv are: a) more local features within the current chunk are leveraged to enhance the performance; b) model weights are fully shared for the two modes to encourage them to promote mutually~\cite{yu2020dual}.

\noindent {\bf 3) Stop-gradient operation.} Empirically, an offline representation would outperform an online one, since the former is conditioned on more future context. Considering that the quantizer will be removed in the downstream fine-tuning stage~\cite{baevski2020wav2vec}, it can be completely turned into offline-mode to enhance the performance during the pre-training. Thus, we apply the stop-gradient operation to $\mathbf q$ obtaining $\widetilde{\mathbf q}$, which decouples the impact of online-mode objectives to the quantizer. Our experiments show that the stop gradient operation largely improves the ASR performance (shown in Sec.~\ref{subsec:discussion}).

\noindent {\bf 4) Joint training loss.} As for the self-supervised pre-training, the contrastive losses~\cite{baevski2020wav2vec} in offline (${\mathcal{L}_{\rm con}^{off}}$) and online (${\mathcal{L}_{\rm con}^{on}}$) modes
are proposed and jointly trained as ${\mathcal{L}}_{\rm con}=\lambda{\mathcal{L}_{\rm con}^{off}}+(1-\lambda){\mathcal{L}_{\rm con}^{on}}$, where $\lambda=0.5$ assigning equal importance for the two items. As for the downstream task, the ASR model is initialized from the pre-trained encoder and then fine-tuned on the labeled dataset $\boldsymbol{D}_l$ in a unified manner same as the pre-training. In this stage, the masking and quantizer of the pre-training are removed. Instead, a fully connected layer is added to the top of the encoder, and a multi-layer Transformer decoder is applied for the ASR task learning~\cite{yao2021wenet}. To train the ASR model, we adopt the representative hybrid loss~\cite{watanabe2017hybrid} ${\mathcal{L}}_{\rm ctc/att}(\mathbf{x}_l,\mathbf{y}_l)={\epsilon}{\mathcal{L}}_{\rm ctc}(\mathbf{x}_l,\mathbf{y}_l)+(1-{\epsilon}){\mathcal{L}}_{\rm att}(\mathbf{x}_l,\mathbf{y}_l)$, with ${\mathcal{L}}_{\rm ctc}(\mathbf{x}_l,\mathbf{y}_l)$ the Connectionist Temporal Classification (CTC) loss and ${\mathcal{L}}_{\rm att}(\mathbf{x}_l,\mathbf{y}_l)$ the attention-based loss, and with ${\epsilon}=0.3$~\cite{yao2021wenet}. Finally, the offline and online hybrid losses (${\mathcal{L}}_{\rm ctc/att}^{off}(\mathbf{x}_l,\mathbf{y}_l)$ and ${\mathcal{L}}_{\rm ctc/att}^{on}(\mathbf{x}_l,\mathbf{y}_l)$) are aggregated for training a unified ASR model as
\vspace{-0.05in}
\begin{equation}
    {{\mathcal{L}}_{\rm asr}}={\alpha}{\mathcal{L}}_{\rm ctc/att}^{off}(\mathbf{x}_l,\mathbf{y}_l)+(1-\alpha){\mathcal{L}}_{\rm ctc/att}^{on}(\mathbf{x}_l,\mathbf{y}_l)
    \label{eq2}
\end{equation}
where $\alpha$ is used to balance the two items (studied in Sec.~\ref{subsec:discussion}).

\vspace{-0.05in}
\section{Experiments and Discussion}
\label{sec:results}
\vspace{-0.1in}
\subsection{Experimental setup}
\label{subsec:exp_setup}
{\bf Data preparation.} The public LibriSpeech (LS)~\cite{panayotov2015librispeech} utterances ($\boldsymbol{D}_u$, 960 hours) and the transcriptions of the train-clean subset ($\boldsymbol{D}_l$, 100 hours) are used in our experiments. The WER performance of the trained models is evaluated on the original test-clean (considered easier) and test-other (considered harder/more noisy) datasets. The 80-dimensional Mel-spectrograms of each speech are pre-processed as the input to the model with 25ms window size and 10ms step size.\\
{\bf Model.} The ASR model consists of 2 Conv sub-sampling layers, 12 Conformer layers for the encoder, and 6 Transformer layers for the decoder~\cite{yao2021wenet}. For the Conv sub-sampling layers, the kernel sizes are (3,3) and the strides are 2. All Conformer and Transformer layers have 8 multi-heads and are 512-dimensional; the kernel sizes of the non-causal Convs are 15. In addition, 5003 modeling units are utilized including 5000 Byte-Pair Encoding (BPE)~\cite{gowda2019multi} units and 3 non-verbal symbols (blank, unknown-unit, and sos-eos). Following~\cite{baevski2020wav2vec}, we set the multiple codebooks in the quantizer with groups 2, entries 320 and vector dimension 512. The number of model parameters is about 0.1 billion, which is comparable with the Wav2vec2-base model~\cite{baevski2020wav2vec}. Both the pre-training and fine-tuning are optimized with a mini-batch of 96 and using the same learning rate scheduler as~\cite{yao2021wenet}.\\
{\bf Decoding.} The models are evaluated with two decoding chunk sizes: 1) $c_s=\infty$ for offline mode; and 2) $c_s=16$ for online mode. The latency of $c_s=16$ is uniformly distributed from 0 to 640ms, with an average latency of 320ms~\cite{yao2021wenet}. The beam size of the CTC Prefix Beam Search (CPBS) decoding~\cite{graves2006connectionist} is 10, and the parameters for the Attention Rescore (AR) decoding are the same with the training~\cite{yao2021wenet}. Note that no language model is applied in the decoding.

\begin{table}[t]
 \vspace{-0.1in}
	\centering  
	\caption{WERs ($\%$) for Causal-Conv (CC) and Chunked-Non-Causal-Conv (CNCC) Conformers in pre-training and fine-tuning.}
	\label{table3}  
	\resizebox{0.9\linewidth}{!}{
	\begin{tabular}{c|c|c|c|c|c|c}  
	    \hline
	    ~&Pre-training&Fine-tuning&\multicolumn{2}{c|}{test clean (AR)}&\multicolumn{2}{c}{test other (AR)}\\
	    \cline{4-7}
	    ~&(offline)&(unified)&offline & online & offline &online \\
	    \hline
	    \hline
		B1&CNCC & CC & 6.78& 7.66&17.65&20.31 \\
		B2&CC & CC & 6.11 & 7.53 & 15.08 & 19.68 \\
		B3&CC & CNCC & 5.90 &7.47&15.04&19.82 \\
		B4&CNCC & CNCC & {\bf 5.62} & {\bf 7.05}& {\bf 13.62}&{\bf 18.11} \\
		\hline
	\end{tabular}
	}
         \vspace{-0.2in}
\end{table}

\vspace{-0.1in}
\subsection{Main experiment}
\vspace{-0.05in}
\label{subsec:main_exp}
{{\bf Baseline methods}\footnote{Note that previous works~\cite{chiu2022self,cao2021improving} are not fully comparable because they are not unified models.~\cite{chiu2022self} optimized two modes separately on a much larger 60k-hour dataset, while~\cite{cao2021improving} distilled an offline model to the online model.}.} We implement two unified baseline models (with 0.1-billion parameters) consist of: 1) a supervised baseline -- Conformer-unified~\cite{yao2021wenet}: a State-Of-The-Art (SOTA) unified ASR system using the conventional causal-Conv Conformer; and 2) an SSL-based baseline -- Wav2vec2-Conformer-unified: we implement the Conformer-unified model while initializing the encoder network from pre-trained Wav2vec2-Conformer~\cite{zhang2020pushing}. 

\noindent{\bf System performance.} As shown in Table~\ref{table1}, we compare UFO2 with SOTA supervised and SSL-based ASR approaches in both offline ($c_s=\infty$) and online ($c_s=16$) modes. As for the systems that are trained for offline uses alone (S1-S3), both the Wav2vec2-base system (S2)~\cite{baevski2020wav2vec} and Wav2vec2-Conformer system (S3)~\cite{zhang2020pushing} largely outperform the supervised Conformer-offline method (S1)~\cite{gulati2020conformer} due to the benefits from SSL on the 960 hours of unlabeled utterances. Compared with S1, the Conformer-unified system (S4)~\cite{yao2021wenet} is compatible with the two modes, while obtaining higher WERs in offline mode due to the use of causal-Conv Conformer and random-mode training. As for the unified ASR method based on SSL, the Wav2vec2-Conformer-unified system (S5) improves the performance compared with S4 via parameters initializing from the pre-trained representation model of S3. Note that since S3 uses non-causal-Conv Conformer, only the left half of the non-causal-Conv kernels are inherited to the downstream S5. Compared with the existing methods, our proposed UFO2 (S6) achieves the best performance with obvious WER reduction in both the two modes. Numerically, compared with the Wav2vec2-Conformer-unified baseline (S5), UFO2 (S6) achieves an average of 29.7\% and 18.2\% relative WER reduction in offline and online modes, respectively.

\vspace{-0.1in}
\subsection{Ablation study}
\label{subsec:ablation}
\vspace{-0.05in}

As shown in Table~\ref{table2}, ablation studies with different training losses and model choices are conducted by comparing our UFO2 (A5) with the Wav2vec2-Conformer-unified baseline (A1) and the variants (A2-A4). First, when comparing A2 with A1, the joint of online and offline ASR losses outperforms the random-mode strategy in the fine-tuning stage. The finding is consistent with the previous study~\cite{yu2020dual}. Compared with A1 that fine-tunes a causal-Conv Conformer initialized from Wav2vec-Conformer~\cite{zhang2020pushing}, A3 uses the chunked non-causal-Conv Conformer (all parameters of Wav2vec-Conformer are inherited), which enhances the ASR performance by leveraging more local features within the chunk. A4 is the combination of A2 and A3, which achieves a further performance improvement. Note that the A4 method can also be extended to 1) other released pre-trained models since most of the existing SSL-based models are pre-trained in an offline manner~\cite{mohamed2022self}, and 2) the supervised setting. 
Finally, UFO2 (A5) combines all of the proposed strategies in both the pre-training and fine-tuning stages, and significantly reduces the WERs compared with the baseline system A1.

\vspace{-0.1in}
\subsection{Discussion}
\label{subsec:discussion}
\vspace{-0.05in}

To further evaluate the effectiveness of UFO2, we analyze the proposed method from the following four perspectives. Due to space limitations, the WER results of the AR decoding are reported for the discussion (We omit the CPBS decoding which is the same trend).

 \begin{table}[t]
 \vspace{-0.2in}
	\centering  
	\caption{An ablation study on the stop gradient in WER ($\%$).}
	\label{table4}  
	\resizebox{0.83\linewidth}{!}{
	\begin{tabular}{l|c|c|c|c}  
	    \hline
	    \multirow{2}{*}{Methods}&\multicolumn{2}{c|}{test clean (AR)}&\multicolumn{2}{c}{test other (AR)}\\
	    \cline{2-5}
	    ~&offline & online & offline &online \\
	    \hline
	    \hline
		UFO2 w/o stop-grad & 5.59 & 6.87 & 13.52 & 17.09 \\
		UFO2 w/ stop-grad & {\bf 4.99} & {\bf 6.35} & {\bf 11.75} & {\bf 16.05} \\
		\hline
	\end{tabular}
	}
        \vspace{-0.2in}
\end{table}

\begin{table}[t]
	\centering  
	\caption{WERs ($\%$) for different fine-tuning hyperparameters.}
	\label{table5}  
	\resizebox{0.87\linewidth}{!}{
	\begin{tabular}{l|c|c|c|c}  
	    \hline
	    \multirow{2}{*}{Weights of offline mode}&\multicolumn{2}{c|}{test clean (AR) }&\multicolumn{2}{c}{test other (AR)}\\
	    \cline{2-5}
	    ~&offline & online & offline &online \\
	    \hline
	    \hline
		$\alpha=0$ & 6.19&6.44&13.81&16.60 \\
		$\alpha=0.25$ &  5.23&6.39&12.59&16.36\\
	    $\alpha=0.5$ &  5.15&{\bf 6.35}&12.05&{\bf 16.04}\\
		$\alpha=0.75$ &  {\bf 4.99}&{\bf 6.35}&{\bf 11.75}&16.05\\
		$\alpha=1$ &  5.13&15.38&12.00&25.67\\
		\hline
	\end{tabular}
	}
        \vspace{-0.2in}
\end{table}

\begin{table}[t]
    \vspace{-0.2in}
	\centering  
	\caption{WERs ($\%$) for Fine-Tuning (FT) on LS-960h.}
	\label{table6}  
	\resizebox{1.0\linewidth}{!}{
	\begin{tabular}{l|c|c|c|c|c}  
		\hline  
		\multirow{3}{*}{Methods}&\multirow{3}{*}{\makecell[c]{Pre-\\training}} & \multicolumn{4}{c}{FT on LS-960h} \\
		\cline{3-6}
		~&~&\multicolumn{2}{c|}{LS-test-clean} & \multicolumn{2}{c}{LS-test-other} \\
		\cline{3-6}
		~&~&offline&online&offline&online\\
		\hline
		\hline
            Conformer-unified~\cite{yao2021wenet} & NA&3.6&4.0	&9.3&11.1\\
            \hline
            Wav2vec2-base~\cite{baevski2020wav2vec} & \multirow{4}{*}{\makecell[c]{LS-\\960h}}&3.4&NA	&8.5&NA\\
            Wav2vec2-Conformer~\cite{zhang2020pushing} & ~&3.1&NA&7.4&	NA\\
            Wav2vec2-Conformer-unified & ~&3.5&4.0&8.5&10.8\\
            UFO2 & ~&{\bf 3.0}&{\bf 3.8}&{\bf 7.1}&{\bf 9.4}\\
		\hline
	\end{tabular}
	}
        \vspace{-0.15in}
\end{table}

\begin{table}[t]
	\centering  
	\caption{WERs ($\%$) for Fine-Tuning (FT) on GS-250h and TL-200h.}
	\label{table7}  
	\resizebox{1.0\linewidth}{!}{
	\begin{tabular}{l|c|c|c|c|c|c|c}
		\hline  
		\multirow{3}{*}{Methods}&\multirow{3}{*}{\makecell[c]{Pre-\\training}} & \multicolumn{4}{c|}{FT on GS-250h}&\multicolumn{2}{c}{FT on TL-200h} \\
		\cline{3-8}
		~&~&\multicolumn{2}{c|}{GS-dev} & \multicolumn{2}{c|}{GS-test} & \multicolumn{2}{c}{TL-test} \\
		\cline{3-8}
		~&~&offline&online&offline&online&offline&online\\
		\hline
		\hline
            Conformer-unified~\cite{yao2021wenet} & NA&24.5&26.1&23.7&25.3&9.6&10.8\\
            \hline
            Wav2vec2-base~\cite{baevski2020wav2vec} & \multirow{4}{*}{\makecell[c]{LS-\\960h}}&20.4&NA	&20.0&NA&8.6&NA\\
            Wav2vec2-Conformer~\cite{zhang2020pushing} & ~&18.5&NA&18.2&	NA&6.7&NA\\
            Wav2vec2-Conformer-unified & ~&21.3&22.6&21.0&22.2&8.3&9.3\\
            UFO2 & ~&{\bf 17.8}&{\bf 20.3}&{\bf 17.4}&{\bf 19.8}&{\bf 6.0}&{\bf 7.7}\\
		\hline
	\end{tabular}
	}
        \vspace{-0.2in}
\end{table}

\noindent{\bf Chunked non-causal Conv in Conformer.} To analyze the effect of the Conv in Conformer blocks, we conduct three experiments (B2-B4) based on the Wav2vec2-Conformer-unified baseline (B1), with causal/chunked non-causal Convs for the pre-training and fine-tuning stages (shown in Table~\ref{table3}). 
Compared with B1, B2 obtains lower WERs with the causal-Conv Conformer in pre-training. We infer that the consistency of model structures between the pre-training and fine-tuning would bring more advantage than only using non-causal Conv for pre-training. The performance of B3 is similar to B2, although a chunked non-causal-Conv Conformer (only the left-half Conv kernels are pre-trained) is adopted for the downstream ASR task in B3. A possible reason is that the chunked non-causal-Conv would bring little improvement if it is not fully pre-trained. Compared with B1-B3, B4 (proposed in UFO2) achieves the best performance via using chunked non-causal-Conv in both the two training stages, which leverages more local features and also fully inherits the pre-trained weights for the downstream fine-tuning.

\noindent{\bf Stop gradient in pre-training.} Inspired by~\cite{chiu2022self}, we decouple the online-mode objectives to the quantizer via stop gradient operation in the pre-training, which largely reduces WERs, as shown in Table~\ref{table4}. The results indicate that the model would learn a high quality quantizer with full-context inputs, while the online-mode objectives might degrade the quantized representation due to a heavy burden on the representation learning when a large proportion of the utterance (i.e. future context) is unavailable~\cite{fu2021scala}.

\noindent{\bf Hyperparameter tuning for fine-tuning.} We analyze the hyperparameter choice for the fine-tuning loss in Eq.~\ref{eq2}, which balances the offline and online terms. As shown in Table~\ref{table5}, $\alpha=0$ and $\alpha=1$ imply fine-tuning an online model and an offline model respectively, which suffer from high WERs when the decoding mode is inconsistent with the training. $\alpha=0.25,0.5,0.75$ imply fine-tuning unified models, which significantly outperform both the online model ($\alpha=0$) and offline model ($\alpha=1$). The results also indicate that our UFO2 achieves the best performance when $\alpha=0.75$. We infer that the online and offline mode promote each other with the joint training manner: the offline model that usually achieves better results would help the online model learning; while the dynamic-chunked strategies might bring advantage like spectrogram augmentation~\cite{park2019specaugment} for the offline model via masking the future context.

\noindent{\bf Fine-tuning on large and out-of-domain datasets.} To further verify the effectiveness of our approach, we fine-tune UFO2 on three different datasets: 1) all of the 960-hour LS dataset, 2) 250-hour GigaSpeech (GS) subset-S for quick research experiments~\cite{chen2021gigaspeech}, and 3) 200-hour TedLium2 (TL) dataset collected from TED talks~\cite{rousseau2014enhancing}. As shown in Table~\ref{table6}-\ref{table7}, UFO2 consistently outperforms the baseline systems, with fine-tuning on large and out-of-domain datasets. Numerically, compared with the SSL-based baseline Wav2vec2-Conformer-unified, UFO2 achieves relative WER reductions by 15.3$\%$/9.0$\%$, 16.8$\%$/10.5$\%$, 27.7$\%$/17.2$\%$ in offline/online modes when fine-tuning on LS-960h, GS-250h, TL-200h, respectively.

\vspace{-0.1in}
\section{Conclusion}
\vspace{-0.1in}
\label{sec:Conclusion}

In this paper, a novel framework named UFO2 is proposed to unify the online and offline ASR models based on SSL, which simplifies the two separate training workflows into a single one, and improves the recognition accuracy. Our results on the LibriSpeech dataset show that the proposed UFO2 significantly enhances the performance compared to the baseline methods. However, we also find that the performance in online mode still underperforms the offline mode, which will be further addressed in the future work.

\vfill\pagebreak

\bibliographystyle{IEEEbib}
\bibliography{strings,refs}

\end{document}